\documentstyle[12pt,aaspp4]{article} 
\tighten

\begin{document}

\title{The Age Dependent Luminosities of the \\
Red Giant Branch Bump, Asymptotic Giant Branch Bump, \\
and Horizontal Branch Red Clump }


\author{ David~R.~Alves\altaffilmark{1} }

\affil{Lawrence Livermore National Laboratory, Livermore, CA 94550 \\
        Email: {\tt  alves@igpp.llnl.gov}}

\altaffiltext{1}{Department of Physics, University of California,
        Davis, CA 95616 }

\author{ Ata~Sarajedini\altaffilmark{2} }
\affil{San Francisco State University, San Francisco, CA 94132 \\
        Email: {\tt  ata@stars.sfsu.edu}}

\altaffiltext{2}{Hubble Fellow}

\begin{abstract}

Color-magnitude diagrams of globular clusters
often exhibit a prominent horizontal branch (HB) and may also
show features such as the red giant branch (RGB) bump and the asymptotic
giant branch (AGB) bump.  
Stellar evolution theory predicts that the luminosities of these
features will depend on the metallicity and {\it age} of the cluster.
We calculate theoretical lines of 2 to 12 Gyr constant age
RGB-bumps and AGB-bumps in the $\Delta$V$^{HB}_{Bump}$--[Fe/H] diagram,
which shows the brightness difference between the bump and the HB as a
function of metallicity.  In order to test the predictions, we identify
giant branch bumps in new {\it Hubble Space Telescope} 
color-magnitude diagrams for 8 SMC clusters.  First, we conclude that 
the SMC cluster bumps are RGB-bumps.   The data for clusters
younger than $\sim$6 Gyr are in fair agreement 
with our predictions for 
the relative age dependent luminosities of the HB and RGB-bump.
The $\Delta$V$^{HB}_{Bump}$ data for clusters older 
than $\sim$6 Gyr demonstrate 
a less satisfactory agreement with our calculations.  We conclude that
$\sim$6 Gyr is lower bound to the age of clusters for which the Galactic
globular cluster, age independent $\Delta$V$^{HB}_{Bump}$--[Fe/H] calibration
is valid.  
Application of the $\Delta$V$^{HB}_{Bump}$--[Fe/H] diagram
to stellar population studies is discussed.

\end{abstract}

\clearpage

\section{ Introduction }

Color-magnitude diagram studies of globular clusters are ideally suited to
test stellar evolution theory (Renzini \& Fusi Pecci 1988).  
The stars in globular clusters are typically coeval and chemically
homogeneous. Thus, the globular clusters represent stellar
populations pure in composition.

A landmark feature in most
color-magnitude diagrams is the horizontal branch (HB).
Horizontal branches are populated by near constant luminosity
core helium-burning giants, which can
exhibit a range of temperatures.   The well known
RR Lyrae variable stars are HB stars with temperatures
inside the boundaries of the instability strip.  Some color-magnitude
diagrams show predominantly red HBs, also known as red clumps
(Cannon 1970), where most of 
the HB stars are cooler than the instability strip.  
Old metal-rich globular clusters
usually have red clumps, as do younger clusters,
which have more massive HB stars.  
We designate all HB stars
cooler than the instability strip, and evolved from red giant branch (RGB) stars
with degenerate helium cores, as {\it red clump 
stars}. Our definition therefore applies to all red horizontal
branch stars in stellar populations older than $\sim$ 0.6 Gyr, i.e.~those
that have undergone the RGB phase transition 
(Sweigart, Greggio, \& Renzini 1990).

Stellar evolution theory predicts that the red HB clump
luminosity depends on age.
When comparing two clusters of similar metallicity,
the red HB clump 
of the younger cluster will generally be brighter
(Lattanzio 1986; Seidel, Demarque, \& Weinberg 1987).
The effect is $\sim$0.3 mag for an age difference of $\sim$10 Gyr\footnote{The 
red HB clump luminosity will actually decrease for very young clusters, i.e. those
just older than the RGB phase transition (e.g Lattanzio 1986).}.
The rate of change
in red clump luminosity with age may also
depend on metallicity (Sarajedini, Lee \& Lee 1995).
Unfortunately,
little observational data exists with which to study
the dependence of red clump luminosity on age, or test for any
dependence at all\footnote{The notable exception is the
Carina dwarf galaxy color-magnitude diagram, which shows a prominent blue (old)
horizontal branch, and a (young) red clump $\sim$0.25
mag brighter (Smecker-Hane et al.~1994).}.  We therefore
seek a new observational test of the theoretically predicted age
dependent red HB clump luminosity in order to
lay the foundation for more
accurate interpretations of color-magnitude diagrams of globular
clusters and field populations in local group galaxies.

In addition to the HB, globular cluster
color-magnitude diagrams may show features such as the RGB-bump,
an evolutionary pause on the first-ascent red giant branch (Thomas 1967; Iben 1968), 
or the AGB-bump, a clustering of stars at the base of the asymptotic giant branch
(Castellani, Chieffi, \& Pulone 1991).
RGB-bumps and AGB-bumps are much more rapid phases of
stellar evolution than the HB, which makes their detection challenging.
However, if an RGB-bump or AGB-bump is identified, 
the brightness difference between the bump
and HB provides a new test of the {\it relative}
age dependent luminosities of these features, free from
uncertainties due to distance or reddening.  

To make our study of the luminosities of the RGB-bump,
AGB-bump, and red HB clump, we first collect the relevant theoretical
evolutionary results and predict the behavior of these features for
clusters of different ages and metallicities.  We endeavour to make
this exercise {\it simple}, and make all of the underlying theoretical
assumptions and calibrations as transparent as possible.  
We then seek a test of
our predictions with a suitable set of observations.  
Although high quality photometric studies of many
Galactic globular clusters have been available for years, these clusters
are predominantly old, and do not span an adequately wide range of ages
(Sarajedini, Chaboyer \& Demarque 1997).
For these reasons, Galactic globulars are a poor place to look. 
On the otherhand, the globular clusters of the 
Small Magellanic Cloud (SMC) range in age from $\sim$2 to 12 Gyr and
are ideally suited to our study.  Color-magnitude diagram photometric data for
clusters in the SMC are best obtained with the high resolution of the
{\it Hubble Space Telescope}.  These new data are
just becoming available (Mighell, Sarajedini \& French 1998).

Our paper is organized as follows.
In \S2 we present a theoretical examination of the age dependencies 
of $M_{V,HB}$, $M_{V,RB}$, and $M_{V,AB}$, the absolute visual magnitudes
of the horizontal branch, RGB-bump, and AGB-bump respectively.
In \S3 we present new {\it Hubble Space Telescope} data for SMC clusters
and review extant detections of RGB-bumps and AGB-bumps in the literature.
In \S4 we use our theoretical data to predict constant age lines in the
$\Delta$V$^{HB}_{Bump}$--[Fe/H] diagram, which shows the brightness
difference between a bump (RGB-bump or AGB-bump) and the HB.   
We then test these predictions with our SMC cluster data.
In \S5, we make our conclusions.

\section{Theoretical Data}

As a wide variety of theoretical data
exist in the literature, we endeavour to make some comparisons
of different studies, which allows a quick estimation of the level
of consistency found.  Wherever possible, we
employ easily reproducible analytic calibrations and simple,
plausible approximations to the theoretical data.  We rely on empirical
calibrations when available.

\subsection{Analytic Calibrations}

We require a calibration between 
turn-off mass and age.  The absolute age scale
is less important than the relative ages.  Following Iben and Laughlin
(1989) and using the theoretical data of Mengel et al.~(1979) we derive,
\begin{equation}
\log(Age) = 10.15   +  0.16 {\rm[Fe/H]}  -  3.86 \log(M) \\
 -  0.2 {\rm[Fe/H]} \log(M)  +  2 \log^2(M)
\end{equation}
where age is in units of years (throughout this paper) 
and $M$ is turn-off mass in solar units.
Equation~1 is applicable for $M \approx 0.7$ to 2.0 $M_{\odot}$ and
[Fe/H] $> -1.3$ dex.  For [Fe/H] $< -1.3$ dex, we substitute
[Fe/H] = $-1.3$ dex, as the age--mass calibration is increasingly
insensitive to metallicity.  
Including metallicity in the age--mass calibration
has only a small effect on the results of this paper.
Equation~1 predicts relative ages for
different mass and metallicity stellar evolution models
with a precision of $\pm$ 1 Gyr.

To an excellent approximation, the bolometric correction ($M_{bol}
= M_V + BC_V$) for giant branch and red horizontal branch stars
with $\log(T)$ = 3.55 to 3.75 is given by,
\begin{equation}
BC_V = -529.226 + 282.524 \log(T) - 37.719 \log^2(T)
\end{equation}
We adopt the usual relation between $M_{bol}$ and $\log(L)$, and
$M_{bol,\odot}$ = 4.69 mag.  Equation~2 is derived from the
Padova isochrones (Bertelli et al. 1994) at metallicities
[Fe/H] = $-0.4$, $-0.7$, $-1.3$, $-1.7$ dex.  We find a negligible
systematic dependence on [Fe/H].  Equation~2 will typically predict
$BC_{V}$ to within 0.01 mag of the Padova data.
To transform the luminosity of an RGB-bump to an observable $M_V$
magnitude, we require $\log(T)$ from an interpolation along reference
RGBs of appropriate age and metallicity.
The reference RGBs have the form,
\begin{equation}
\log(L) = \alpha + \beta \log(T) + \gamma\log^2(T) + f(Age)
\end{equation}
\begin{equation}
f(Age) = -0.502\log(Age) - 4.168
\end{equation}
where the coefficients $\alpha$, $\beta$ and $\gamma$ depend on
metallicity and are given in Table~1.
These are also derived
from the Padova isochrones, valid in the luminosity range $\log(L)
= 1.3$ to 3.0, and ages of 2 to 15 Gyr.  These give $\log(T)$ along
the giant branch with an accuracy better than $\delta\log(T)$ = 0.005,
which corresponds to $\pm 0.05$ mag in $M_V$ using Equation~2.  

\subsection{Horizontal Branch}

The dependence of the absolute visual magnitude of the HB on metallicity
has been the subject of numerous theoretical and observational studies.
Without further discussion, we will begin
with an empirical $M_{V,HB}$--[Fe/H] calibration, which is applicable
to ancient globular clusters, assuming their ages are the same to within
a few Gyr.  We adopt,
\begin{equation}
M_{V,HB} = 0.15{\rm[Fe/H]} + 0.70
\end{equation}
from Walker (1992), which is consistent with that found by
Alcock et al. (1997) from LMC RR Lyrae stars.  The zero-point
of Equation~5 only enters our analysis
as a negligible second order effect through the bolometric corrections.

In Figure 1, we plot $M_{V,HB}$ as a function of $\log(Age)$, comparing
theoretical data from several different authors; the different data are
described in the caption.  We have used Equations~1 \& 2 to calculate 
$M_{V,HB}$ and age for the data from 
Seidel, Demarque, and Weinberg\footnote{These theoretical data are HB stellar
evolution models.  We have assumed no mass-loss on the first-ascent RGB, and
equate total HB model mass with turn-off mass.  This assumption may be poor
for the 0.9 and 1.0 mass models, but is likely satisfactory for the higher
mass (younger) models. See also discussion in \S2.4 of this paper.}
(1987), Vassiliadis and Wood (1993), and Lattanzio (1986).  The data of
Sarajedini, Lee and Lee (1995) were presented as $M_{V,HB}$ and age,
and we have adopted them as published.
The data of Sarajedini, Lee and Lee (1995) suggest that the dependence of $M_{V,HB}$ 
on age may be a function of metallicity.  The data of Vassiliadis and
Wood (1993) do not show such a metallicity dependence, although the limited number
of data-points precludes any strong statements to this effect.   
We will simply adopt the following relation for all metallicities, ignoring
a possible age-metallicity cross-term,
\begin{equation}
M_{V,HB} \propto 0.5\log(Age)
\end{equation}
which is shown in the lower right corner of Figure~1 and labeled.
As a check of Equation~6, consider the
dual horizontal branches of the Carina dwarf (Smecker-Hane et al.~1994;
see also Hesser et al. 1996).  Carina
is $\sim$80\% of a 2.5 to 7 Gyr old population and $\sim$20\% of 10 to 14 Gyr old
population; both components have similar metallicities. 
Assuming the red clump represents a population $\sim$4 Gyr old,
and the blue HB represents a population $\sim$12 Gyr old,
Equation~6 predicts a brightness difference of 0.24 mag,
in good agreement with 0.25 mag observed.

\subsection{RGB-Bump}

The RGB-bump is theoretically understood as a luminosity dip 
due to the hydrogen-burning shell crossing a discontinuity in chemical
composition left by the deepest penetration of the convective envelope
(Thomas 1967; Iben 1968).
The discontinuity represents a drop in the mean molecular weight of fuel
and an increased opacity for the hydrogen-burning shell 
as it moves radially outward, which causes a drop in
luminosity and small increase in temperature (Refsdale \& Weigert 1968;
Sweigart, Greggio \& Renzini 1990).  The RGB-bump is predicted for 
a wide range of masses, and metal and helium abundances (Sweigart \& Gross   
1978; Sweigart, Greggio \& Renzini 1989), and is insensitive to
``smoothing'' of the composition discontinuity due to convective undershooting
(Bono \& Castellani 1992).
Alongi et al. (1993) showed that the dependence of RGB-bump luminosity
on mass is independent of overshoot or classical mixing schemes.
The number of stars decreases
with increasing luminosity along the RGB (Castellani, Chieffi \& Norci 1989),
which makes observational detection of the RGB-bump challenging for
low metallicity or young populations (these RGB-bumps would be bright).  The difficulty 
is compounded in young populations, as the duration of the pause is shorter
relative to the pace of evolution up the RGB,
and thus the contrast of RGB-bump stars against the underlying RGB 
luminosity function is less.

As noted by Fusi Pecci et al.~(1990), theory predicts absolute
RGB-bump luminosities brighter than observed.  However, they found
good agreement between theory and observation for the
metallicity dependence of the RGB-bump luminosity. 
Cassisi and Salaris (1997) found that updated input physics
resolves the discrepancy between the
predicted and observed absolute luminosities of RGB-bumps.
We will adopt the dependence of RGB-bump luminosity on
mass from theory, but empirically calibrate the absolute luminosities
to globular cluster data.  Fortunately, the number of observational detections
of RGB-bumps for Galactic globular clusters has grown
considerably since the first detection by King et al.~(1985) for 47~Tuc.
In an important paper, Fusi Pecci et al.~(1990) identified the RGB-bump
in 11 Galactic globular cluster luminosity functions.
Sarajedini and Forrester (1995) added data for 5 additional clusters and found,
\begin{equation}
{\rm[Fe/H]} = -1.33 + 1.43\Delta V^{HB}_{Bump} 
\end{equation}
where $\sigma_{RMS}$ = 0.04 dex about this fit.  $\Delta$V$^{HB}_{Bump}$
is defined as the brightness difference, $\Delta$V(Bump$-$HB), between the RGB-bump
and the HB (see \S4 and Fig.~5 of this paper).  
From Equations 5 \& 7, applicable to ancient globular clusters, we find,
\begin{equation}
M_{V,RB} = 0.85{\rm[Fe/H]} + 1.63
\end{equation}
which yields $M_{V,RB}$ = 0.185, 0.525, 1.035, and 1.290 mag for
[Fe/H] = $-1.7$, $-1.3$, $-0.7$, and $-0.4$ dex respectively, our
normalization points.  We are using the subscript ``RB'' to designate
the RGB-bump.  To maintain
consistency with the zero-point in Equation~5, we will assume
that all of the Galactic globular
clusters are 12 Gyrs old (Chaboyer, Demarque \& Sarajedini 1996;
see their Table~3, Column~5),
which yields turn-off masses $M$ = 0.865, 0.865,
0.920, and 0.947 $M_{\odot}$ for the normalization points as above,
from Equation~1.

In Figure~2 we plot $\log(L)$ of the RGB-bump as a function of model
mass, which we will equate with turn-off mass.  The theoretical data 
from Alongi et al.~(1993), Sweigart and Gross (1978), Vassiliadis and Wood (1993),
and Fusi~Pecci et al.~(1990) illustrate the various dependencies of RGB-bump
luminosity on Z, Y, and mixing schemes, and are described more completely
in the caption.  In all of the data, the dependence
of $\log(L)$ on mass is quite similar.  
The solid line in the lower right corner 
of Figure~2 (labeled) represents the dependence
of RGB-bump luminosity on mass adopted in this paper,
\begin{equation}
\log(L) \propto 0.75 M
\end{equation}
The zero-point is arbitrary.  
We now calculate $M$ for ages
of 10, 5, and 2 Gyr and $\Delta\log(L)$ for the RGB-bump
using $\Delta\log(L)_{Age}$ = 0.75$\times$($M_{Age}$ $-$ $M_{12}$), which
follows from Equation~9.  
We calculate $\log(L)$ for the 12 Gyr RGB-bumps from the values of
$M_{V,RB}$ given above (Eqn.~8) using the appropriate reference RGBs.
These values, coupled with $\Delta\log(L)$ as above,
yields $\log(L)$ for the younger RGB-bumps,
calibrated to the Galactic globular cluster data.
Interpolating along the appropriate reference
RGB gives $\log(T)$, $BC_V$, and $M_{V,RB}$.

Our calculations show the dependence of $M_{V,RB}$ on
age varies some with metallicity.  For easy comparison
with Equation~6, which gives the age dependence of $M_{V,HB}$, we
present the following approximation to the different metallicity
$M_{V,RB}$ data,
\begin{equation}
M_{V,RB} \propto 1.7\log(Age)
\end{equation}
Equation~10 is a reasonable approximation because we have
adopted a single mass-luminosity calibration for all metallicities,
i.e.~there is no $M\times$[Fe/H] term in Equation~9.
We conclude that bolometric corrections and the shape of the different
age and metallicity RGBs (in the $\log(L)$--$\log(T)$ plane) have only
a small effect on the age dependence of $M_{V,RB}$.
In any case, $M_{V,RB}$
increases more rapidly with age than $M_{V,HB}$\footnote{Straniero and
Chieffi (1991) give $M_{V,RB} \propto 0.9562\log(Age)$, applicable to
globular clusters with ages from 10 to 20 Gyr.  We speculate that the shallower
dependence on age derives from consideration of theoretical data constructed for
only very ancient clusters.  It may indicate a more complex age dependence than the
simple proportionality given in Eqn.~10.}.

Equations 6 \& 10, which give the age dependencies of $M_{V,HB}$ and $M_{V,RB}$,
illustrate a fundamental point of this paper.  Theory predicts the age dependencies
of the red HB clump and RGB-bump luminosities to be sufficiently different,
such that, in young clusters, the brightness difference $\Delta$V$^{HB}_{Bump}$
will be observably ``shifted'' off the Galactic globular cluster
$\Delta$V$^{HB}_{Bump}$--[Fe/H] calibration (Eqn.~7).
We emphasize that theoretical uncertainties associated with mass-loss on the RGB, 
the reference RGBs, the turn-off mass/age calibration, and the absolute
luminosities of the RGB-bump and HB are second order effects
because we will restrict ourselves to an empirically calibrated,
differential comparison of $M_{V,HB}$ and $M_{V,RB}$.

\subsection{AGB-Bump}

The AGB-bump occurs 
at the beginning of helium shell-burning asymptotic
giant branch (AGB) evolution.  
Caputo et al.~(1989) found the AGB-bump approaches the Hayashi
line in higher mass models.  From an observational perspective,
particularly in young clusters, there is potential for confusion
between the RGB-bump and AGB-bump.  
Theory tells us that the luminosity difference 
between the AGB-bump and HB is fairly insensitive to metallicity
(Castellani, Chieffi \& Pulone 1991)
or helium abundance (Bono et al.~1995).
Theoretical treatment of central helium-burning convective core instabilities
known as breathing pulses\footnote{Modeling the transition
from core helium-burning to helium shell-burning is a matter of considerable debate, 
with implications for the measurement of the primordial helium content,
a parameter of strong cosmological relevance (Caputo et al. 1989).}
have a $\sim$0.25 mag effect on the luminosity
of the AGB-bump (Castellani et al.~1985; Caputo et al.~1989; Bono et al.~1995).
Breathing pulses have been {\it suppressed} in the stellar evolution
models examined in this paper.
Pulone (1992) and Ferraro (1992) both claim the AGB-bump
is a standard candle, with $M_{V,AB} = -0.4 \pm 0.1$ mag for a limited
range of metallicities (low) and ages (ancient).  
There is little discussion of the age dependence 
of the AGB-bump luminosity in the literature. 

In contradistinction to our empirical calibrations of
$M_{V,HB}$ and $M_{V,RB}$, we will predict $M_{V,AB}$ as a function
of age and metallicity directly from theory. 
In a classical paper, Castellani, 
Chieffi, and Pulone (1991) presented a grid of horizontal branch stellar 
evolution models, intended to represent Galactic globular clusters.
The models were evolved 
through to the re-ignition of the hydrogen
shell, i.e.~the beginning of double shell-burning and the putative
onset of the thermal-pulsing AGB phase.  Following standard
procedure, for each metallicity, the mass of the helium core was fixed
according to theoretical prescriptions of the helium flash at the tip of
first ascent RGB ($M_{Core} \approx 0.5 M_{\odot}$).  Total HB mass was
treated as a free parameter, covering a range 
from 0.80 to 0.525 $M_{\odot}$.  To convert HB model mass to turn-off mass, 
we must account for
RGB mass-loss.  We adopt 0.10~$M_{\odot}$ lost on the 
RGB\footnote{ See Figure 8 of Sweigart, Greggio, and Renzini (1990).  Our adopted
RGB mass-loss is roughly equivalent to a Reimers scaling of $\eta$ = 1/3.}.
To extend the theoretical data
to higher masses, we use the models of Vassiliadis and Wood (1993).

In Figure~3, top panel, 
we plot $\log(L)$ of the AGB-bump as a function of
turn-off mass for Z=0.001.
The Castellani 
et al.~data are shown as open squares.  The Vassiliadis and Wood data are
shown as filled circles.  In Figure~3, bottom panel, we plot $\log(L)$ of the AGB-bump
as a function of [Fe/H].  Symbols are the same as in the
top panel.  The Castellani et al.~data are for a turn-off mass of 0.85 $M_{\odot}$, 
while the Vassiliadis and Wood data are for 1.50 $M_{\odot}$.
We adopt the following analytic fit to these data,
\begin{equation}
\log(L) = 1.90 + 0.215(M) - 0.155(M){\rm[Fe/H]}
- 0.047{\rm[Fe/H]}^2
\end{equation}
In the top panel, we project lines from Equation~11 for [Fe/H] =
$-0.4$, $-1.3$, and $-1.7$ dex; each line is labeled.  In the bottom
panel we project lines from Equation~11 for turn-off masses 0.85, 0.95, 1.10,
1.30 and 1.50 $M_{\odot}$; each line is labeled.  
In the absence of more theoretical
data, Equation~11 is a plausible extrapolation of the
Castellani et al.~data to higher masses.  Clearly, more theoretical
models are needed.
To calculate absolute $M_{V,AB}$ we assume
the AGBs are uniformly hotter than the appropriate age and metallicity
reference RGBs by $\delta\log(T)
= 0.02$.  

We find $M_{V,AB}$ increases with age more
rapidly for low metallicities, which is a direct result of the
$M\times$[Fe/H] cross-term included in Equation~11.  In all cases,
the dependence of $M_{V,AB}$ on age is steeper than that 
for $M_{V,HB}$ (Eqn.~6) and shallower than that for $M_{V,RB}$
(Eqn.~10).  Ignoring the metallicity dependence and adopting the mean slope 
of linear regressions of the $M_{V,AB}$ data yields the following approximation,
\begin{equation}
M_{V,AB} \propto 0.90\log(Age)
\end{equation}
where the slope ranges from $\sim$0.8 for the high metallicity AGB-bumps
to $\sim$1.0 for the low metallicity AGB-bumps.  
We predict $M_{V,AB}$ will always be brighter
in the younger of two stellar populations with similar metallicity,
which allows unambiguous identification of RGB-bumps
in certain regions of the $\Delta$V$^{HB}_{Bump}$--[Fe/H] diagram.
For example, $\Delta$V$^{HB}_{Bump} > -0.8$ mag is likely to be an RGB-bump,
regardless of the age or metallicity of the stellar population.  If 
$\Delta$V$^{HB}_{Bump} < -0.8$ mag, the situation is less clear.
The bump maybe a young RGB-bump or an AGB-bump.

\subsection{Summary of Theoretical Data}

Table~2 summarizes our theoretical calculations for the
absolute visual magnitudes of
the red HB clump, RGB-bump, and AGB-bump.
Column~1 lists [Fe/H] in units of dex.  Column~2 gives age in Gyr.  
Column~3 gives turn-off mass calculated from the age according to Equation~1.
Column~4 gives $M_{V,HB}$ in units of magnitudes
for our 12 Gyr old calibrators according to Equation~5, and for 10, 5, and 2 Gyr
assuming an age dependence given by Equation~6, relative to the 12 Gyr old
calibrators.  Columns~5 \& 6 list the values 
of $M_{V,RB}$ and $M_{V,AB}$ respectively.
Figure~4 illustrates the dependencies
on age for $M_{V,HB}$, $M_{V,RB}$, and $M_{V,AB}$ given
Equations~6, 10 \& 12 of this paper respectively, taken
directly from Table~2.  We plot $M_V$ as a function of $\log(Age)$;
each of the four panels represents a different metallicity,
which is labeled.  The HB data are shown as
filled squares, the RGB-bump data as open circles, and the AGB-bump
data as open triangles.

\section{Observational Data}

In order to test the predictions of the theoretical calculations, we must
compare them to observational data. In this case, we require information
on the ages, metallicities and $\Delta$V$^{HB}_{Bump}$ values for several
clusters. The ideal dataset for this purpose is that of Mighell et al.~(1998),
where {\it Hubble Space Telescope} (HST) observations are combined with
ground-based photometry to estimate the ages and abundances of 7 star
clusters in the SMC (Lindsay 113, Kron 3, NGC 339,
NGC 416, NGC 361, Lindsay 1, and NGC 121). We refer the reader to that
paper for a description of how these quantities were measured. 
Note that in the present work, the uncertainties quoted for
the V(HB) values are the standard error of the mean (random error), 
whereas in the paper by Mighell et al. (1998), the V(HB) errors 
represent the total uncertainty (random as well as systematic). This is done
because we are primarily interested in the {\it difference} in magnitude between
the bump and the HB, not the aboslute value of each quantity.
To establish
values for the magnitude of the bump in each cluster, we proceed as follows.
First, we construct a luminosity function (LF) of the giant branch. We then perform a
Gaussian fit to the LF in the region of the bump. The peak of
this fit is the value of $V(Bump)$. The $\sigma$ of the
Gaussian divided by the square-root of the number of points used in the fit
(i.e. the area under the Gaussian) is then our estimate for the
error in $V(Bump)$. 

In order to supplement this dataset with clusters of even younger age, 
we searched the HST archive and constructed color-magnitude diagrams
for approximately two dozen populous LMC and SMC clusters that have Wide Field
Planetary Camera 2 (WFPC2) observations. This search yielded one SMC cluster,
NGC~411, that exhibits a noticeable clustering of stars on its 
giant branch. The photometry has been reduced following the procedure
outlined by Sarajedini (1998).  The HST color-magnitude diagram of NGC~411,
showing only the data from the Planetary Camera, which was centered on the
cluster, is shown in Figure~5.
We plot $V$ as a function of $(B-V)$; the red HB clump and
giant branch bump are marked with two arrows.
We note that the giant branch bump
$looks$ like it lies on the first ascent RGB, and not along an AGB.

In measuring the age, abundance, and $\Delta$V$^{HB}_{Bump}$ of NGC~411, we
follow procedures consistent with those of Mighell et al.~(1998, see above).
To determine the red HB clump magnitude of NGC~411, we begin by selecting 
stars in the range $0.56 < (B-V) < 0.81$ and $18.85 < V < 19.90$; we find
$V(HB)=19.43\pm0.01$, where the quoted error
is the standard error of the mean.  The V magnitude of the bump
in NGC~411 is measured in the manner as described above.
The metallicity of NGC~411 can be 
estimated by employing the simultaneous reddening and metallicity
(SRM) method of Sarajedini (1994). We note, however, that since NGC~411
appears to be significantly younger than the clusters considered by
Sarajedini (1994) and Mighell et al.~(1998), we must employ a different 
relation between
metallicity and RGB position/shape.  Recently, Noriega-Mendoza \& 
Ruelas-Mayorga (1997) have presented such relations derived from
photometric observations of Galactic open clusters, which nicely encompass the
age and likely abundance of NGC~411. When coupled
with a polynomial describing the shape and location of the NGC~411 RGB, the 
SRM method then yields the metallicity that we seek, 
[Fe/H] = $-0.68\pm0.07$. 
Finally, an estimate of the age of NGC~411 can be derived by appealing
to the isochrones of Bertelli et al. (1994). Utilizing the tracks for
Z = 0.004 and offsetting the isochrones to match the
magnitude of the red HB clump and the color of the main sequence, we find 
an age between 1.26  and 1.58 Gyr for
NGC~411. This is in excellent agreement
with the results of Da Costa \& Mould (1986) who conclude that the age is
$1.5 \pm 0.5$ Gyr. For the present paper, we will adopt $1.4\pm0.2$ Gyr.

Table~3 lists the relevant observational quantities for these 8 SMC
clusters.  Column~1 is cluster name, Columns 2 \& 3 are $V(HB)$
and $V(Bump)$ respectively, in units of magnitudes.  Column~4 is age
in units of Gyr.  Table~3 also summarizes the Galactic
globular cluster RGB-bump data
assembled from Fusi Pecci et al.~(1990) and Sarajedini and Forrester 
(1995).  Where available, we have listed ages from
Chaboyer, Demarque, and Sarajedini (1996), appropriate for the
zero-point adopted in Equation~5 of this paper.

Observational detections of AGB-bumps in local group field population
and globular cluster color-magnitude diagrams have been discussed recently
by Gallart (1998).  For Galactic globular clusters,
we find four $marginal$ detections of AGB-bumps in the literature.
From the luminosity functions presented by Ferraro (1992; see also their Table~2),
we find $\Delta$V$^{HB}_{Bump} = -1.0$, $-1.0$, and $-0.95$ mag for
the AGB-bumps in M5, NGC~1261, and NGC~2808 respectively.  From
the color-magnitude diagram of Hesser et al.~(1987) and the discussion
given by Castellani et al.~(1991) of this data, we adopt $\Delta$V$^{HB}_{Bump}
= -0.95$ for 47~Tuc.  We adopt a metallicity of [Fe/H] = $-1.40$ dex for M5
following Ferraro (1992).  The metallicities for
NGC~1261, NGC~2808 and 47~Tuc are given in Table~3 (47~Tuc = NGC~104).
We caution that $\Delta$V$^{HB}_{Bump}$ for 47~Tuc 
is determined {\it by eye}, following Castellani et al.~(1991, their Fig.~10),
and should be regarded with proper skepticism.
For M5, NGC~1261, and NGC~2808 we estimate an uncertainty of $\pm 0.07$ mag
for each value $\Delta$V$^{HB}_{Bump}$
by inspection of the Ferraro (1992) luminosity functions.
We note that Caputo et al.~(1989) derive 
$\Delta$V$^{HB}_{Bump} = -0.95 \pm 0.09$ mag for the M5 AGB-bump,
in agreement with our estimation.

\section{The $\bf{\Delta}$V$^{\bf{HB}}_{\bf{Bump}}$--[Fe/H] Diagram}

In Figure 6,
we present the theoretical  $\Delta$V$^{HB}_{Bump}$--[Fe/H] diagram,
which shows the brightness difference between
the bump (AGB-bump or RGB-bump) and the HB.
The observational data for RGB-bumps in Galactic globular clusters
are shown as solid black squares.
Error bars are omitted for clarity, but see Table~3 of this paper.
The solid line through these points, and labeled ``12 Gyr'' is the best
fit regression of these points, Equation~7 of this paper.
Assuming the horizontal branch has $no$ dependence on age, the
locations of RGB-bumps in stellar populations of ages 10, 5, and 2 Gyr
are shown as solid lines, and labeled (R10, R5, and R2 respectively).
Here, we are adopting the 12 Gyr
$M_{V,HB}$ values from Table~2 {\it for all ages}.
The short-dashed lines show the
locations of RGB-bumps assuming $M_{V,HB}$ depends on age as given
by Equation~6, and listed in Table~2.  The effect of an age dependent
$M_{V,HB}$ is to move the young RGB-bumps closer to the ancient cluster
calibration in the $\Delta$V$^{HB}_{Bump}$--[Fe/H] diagram.
At ages of $\sim$5 Gyr or younger, we should clearly see
the age-shift effect in observational data.
The dotted lines, labeled A10/12, A5, and A2, show the locations
of AGB-bumps in this diagram, assuming the age-dependent values
of $M_{V,HB}$, as listed in Table~2. The 12 and 10 Gyr lines are
virtually indistinguishable.
The 5 and 2 Gyr old AGB-bump lines
have increasingly more negative $\Delta$V$^{HB}_{Bump}$ values.
The observational detections of AGB-bumps in Galactic globular
clusters, as discussed in \S3, are plotted as filled circles.

Several aspects of Figure~6 warrant further discussion.  We have
plotted the RGB-bump constant age lines under two assumptions of
the dependence of $M_{V,HB}$ on age, the case of $no$ dependence
and that given by theory (Eqn.~6), to illustrate the discriminating
power of RGB-bumps in young clusters.  Assuming the age dependence
of $M_{V,RB}$ is accurately predicted by theory, this is a new test
of the age dependence of $M_{V,HB}$.   In reality, we have no empirical
data with which to constrain the age dependencies of either $M_{V,HB}$
or $M_{V,RB}$, and we will only test the $relative$ age dependencies.  
However, empirical confirmation of the relative age dependencies
of $M_{V,RB}$ and $M_{V,HB}$ is an important new test of our
theories.  As for the AGB-bumps, our theoretical A10/12 line
is in excellent agreement with the detections in Galactic globular
clusters.  While we do not wish to make strong conclusions from this
small and inhomogeneous observational dataset, the
agreement implies the bright zero-point in our $M_{V,HB}$--[Fe/H] calibration
(Eqn.~5) is correct.  An increase in the zero-point of $\sim$0.3 mag, as
advocated by, for example, Carney et al.~(1992), would
move the A10/12 line 0.3 mag to the left in Figure~6, and the observed AGB-bumps
would be too faint.  In turn, if breathing pulses had not been suppressed
in the stellar evolution models examined in this paper, the theoretical
A10/12 line would move $\sim$0.25 mag to the right, in better accord with a
fainter absolute calibration of $M_{V,HB}$.
More theoretical studies of the AGB-bump and more robust
observational detections of AGB-bumps in Galactic globular
clusters would be desirable.

Also shown in Figure~6 are the data for the 8 new SMC cluster
giant branch bumps,
shown as open circles with error bars.
We find four of the SMC clusters
(NGC 416, NGC 361, Lindsay 1, and NGC 121; mean age = 9.0 Gyr) 
lie right on the Galactic
globular cluster RGB-bump calibration.  Three of the clusters
(Lindsay 113, Kron 3, NGC 339; mean age = 5.9 Gyr) are shifted ``off'' of the Galactic
globular cluster RGB-bump calibration by $\sim$2-3$\sigma$ in $\Delta$V$^{HB}_{Bump}$
and $\sim$1-2$\sigma$ in [Fe/H].  
The youngest SMC cluster (NGC~411; age = 1.4 Gyr) deviates
from the Galactic globular cluster RGB-bump calibration with an extremely
high significance.  Moreover, it lies $\sim$8$\sigma$ away from the A10/12
AGB-bump line and $\sim$16$\sigma$ from the A2 AGB-bump line, where the latter
is the most appropriate comparison given the age of NGC~411.  
We conclude that {\it all 8 SMC cluster bumps are RGB-bumps}.
It is also clear that the SMC cluster data, 
particularly NGC~411, are in better agreement
with interpolations between the R2, R5, and R10 lines calculated
with an age dependent $M_{V,HB}$ (the dashed lines) as opposed to
an age independent $M_{V,HB}$ (the solid lines).

In Figure~7, we plot the deviation from the Galactic globular cluster
RGB-bump line in the $\Delta$V$^{HB}_{Bump}$--[Fe/H] diagram
($\delta\Delta$V$^{HB}_{Bump}$) as a function of age, in Gyr.
Plotted as open circles with error bars are the SMC cluster
data from Table~3.  In the lower portion of the figure, underneath
each SMC cluster datapoint, we have labeled the metallicity for each
cluster.  We have plotted the Galactic globular cluster RGB-bump data
as filled squares.  Error bars are omitted for clarity.  Clusters
NGC~5927 and NGC~6637 are not plotted (see Table~3).
The case of $\delta\Delta$V$^{HB}_{Bump}$ = 0 is plotted
as a solid line.  Shown as dotted lines are the theoretical predictions
calculated in this paper for [Fe/H] = $-1.7$, $-1.3$, $-0.7$, and $-0.4$ dex,
using the data in Table~2.
The [Fe/H] = $-1.7$ line deviates from the other three metallicity lines, which
lie close together.  Shown as a dashed line is the approximation to our
theoretical calculations derived from
Equations 6 \& 10 of this paper, which yield 
$\delta\Delta$V$^{HB}_{Bump}$ $\propto$ $1.2\log(Age)$.  
For the SMC cluster younger than $\sim$6 Gyr, we find reasonable agreement
with the theoretical calculations in this paper, lending support
to our conclusion that these giant branch bumps, particularly the
NGC~411 bump, are RGB-bumps.  
These data demonstrate, for the first time, 
deviations from the $\Delta$V$^{HB}_{Bump}$--[Fe/H]
Galactic globular cluster calibration, which are in fair agreement with the
theoretically predicted age dependence of the luminosities of the
RGB-bump and red HB clump.
Interestingly, the data appear to show a 
``break'' near 6 Gyr, where the clusters older than this 
follow the $\Delta$V$^{HB}_{Bump}$--[Fe/H] Galactic globular cluster 
calibration remarkably well.   

\section{Conclusions}

We have examined a variety of theoretical
data regarding the age and metallicity dependent luminosities of the 
red HB clump, the RGB-bump, and AGB-bump.  We make no claim to have provided
an exhaustive review.  Rather, we have endeavoured to make $some$ comparisons
of different theoretical studies, which allows a quick estimation of the
level of consistency found.
We then adopt the simple, plausible approximations to the theoretical
data, such as our
linear $M_{V,HB}$--$\log(Age)$ calibration (Eqn.~6) or the linear $log(L)$--$M$
calibration for the RGB-bump luminosity (Eqn.~9).  
Transformation of the theoretical data
to the observable plane (i.e.~visual magnitudes) is accomplished with 
analytic formulae, which can be easily reproduced.  Where possible, we have
empirically calibrated our calculations; employing, 
for example, the $M_{V,HB}$--[Fe/H]
calibration (Eqn.~5) and the $\Delta$V$^{HB}_{Bump}$--[Fe/H] calibration
(Eqn.~7).  As for the AGB-bump, we have appealed to the models of
Castellani et al.~(1991) and Vassiliadis and Wood (1993) to derive
an analytic approximation for the luminosity as a function of
mass and metallicity (Eqn.~11).  Our final product is
a theoretical $\Delta$V$^{HB}_{Bump}$--[Fe/H] diagram,
showing the brightness difference between the bump
(RGB-bump or AGB-bump) and HB as a function of {\it age} and metallicity.
Our diagram shows good agreement
with extant observational data on RGB-bumps and AGB-bumps
in ancient Galactic globular clusters. 

To test the (younger) constant age RGB-bump and AGB-bump lines
in the $\Delta$V$^{HB}_{Bump}$--[Fe/H] diagram, we have identified
giant branch bumps in the {\it Hubble Space Telescope} color-magnitude diagrams
(Mighell et al.~1998) of 7 SMC globular clusters ranging in age from $\sim$5 to 12 Gyr.
To supplement these data, we have reduced archival {\it Hubble Space Telescope}
data for the SMC cluster NGC~411, which also shows a giant
branch bump.   As the ages and metallicities of these 8 SMC clusters
are constrained by other techniques, we conclude from their locations
in the $\Delta$V$^{HB}_{Bump}$--[Fe/H] diagram that these bumps are
RGB-bumps.  We make three principal conclusions from our comparison
of the observational data and theoretically predicted values
of $\Delta$V$^{HB}_{Bump}$, 
(1) Deviations from the 
Galactic globular cluster $\Delta$V$^{HB}_{Bump}$--[Fe/H]
calibration are consistent with a general
trend of increasing RGB-bump brightness with decreasing age, in
agreement with the theoretical prediction.
(2) The data are in poor agreement with the hypothesis of $no$
dependence on age for the absolute visual magnitude of the red HB clump.
Assuming the age
dependent luminosity of the RGB-bump is accurately predicted by
theory, the red HB clump becomes brighter with decreasing age. 
(3) The data for clusters older than $\sim$6 Gyr follow the
empirical, age independent, Galactic globular cluster 
$\Delta$V$^{HB}_{Bump}$--[Fe/H] calibration (Eqn.~7) well,
which indicates a lower bound to the age of clusters for which this
calibration is applicable.  The calibration would appear to be
an accurate metallicity diagnostic
($\sigma_{Fe/H} = 0.04$ dex; Sarajedini \& Forrester 1995) for clusters
older than $\sim$6 Gyr.

The SMC cluster data show the
worst agreement with our predictions
near ages of $\sim$6 to 10 Gyr, which we have
designated as the 6 Gyr break.  
The 6 Gyr break likely reflects an over-simplification in our
construction of the theoretical $\Delta$V$^{HB}_{Bump}$--[Fe/H] diagram.  
One explanation would be a more complex
dependence of either $M_{V,HB}$ or $M_{V,RB}$ on age 
than employed here.  In this case, the relative
age dependencies of the red HB clump and RGB-bump luminosities likely
approaches our calculation for young stellar populations. 
The luminosities of the red HB clump and RGB-bump
for more ancient populations are not necessarily independent of age,
but likely have a similar dependence on age.
Alternatively, we note the effects of
helium abundance (e.g.~Catelan \& De Freitas Pacheco 1996)
have been neglected in our analysis.   A detailed
accounting for the dependence of $M_{V,HB}$ and $M_{V,RB}$ on helium
abundance is beyond the scope of our paper.
We have assumed possible differences in the helium abundances
of the clusters observed are small.
The observational dataset is still limited.  The identification
of RGB-bumps in other globular cluster color-magnitude diagrams, 
in particular those ranging in age 
from $\sim$1 to 10 Gyr would be highly desirable.

The $\Delta$V$^{HB}_{Bump}$--[Fe/H] diagram may be used 
as a consistency check for age and metallicity determinations measured
via other techniques for globular clusters.  
In particular, $\Delta$V$^{HB}_{Bump}$
is shown to be an accurate predictor of metallicity for clusters as young as
$\sim$6 Gyr, which likely encompasses the ages of all
Galactic globular clusters.  
Additionally, the $\Delta$V$^{HB}_{Bump}$--[Fe/H] diagram may be
employed as a diagnostic tool to aid in the interpretation of
color-magnitude diagrams
of field populations in nearby galaxies, where the different
ages and metallicities of the component stellar populations
are not necessarily known with high accuracy.  In this case,
giant branch bumps may be identified as either RGB-bumps or AGB-bumps
by their locations in the $\Delta$V$^{HB}_{Bump}$--[Fe/H] diagram.
Moreover, the bumps may yield new clues to the component
stellar populations of local group galaxies, which has important implications 
for our theories of galaxy formation.
An excellent example of this would be the 9 million star color-magnitude
diagram of the LMC bar, assembled from the MACHO Project's microlensing
survey photometry (Alves et al.~1998, 1998b; Alcock et al.~1998),
where the very large number of stars will make 
giant branch bump detections statistically very significant.

\section{Acknowledgments}

We thank Peter Wood for 
kindly providing his stellar evolution models
in electronic form.  We thank our referee, Vittorio Castellani,
for his helpful comments and criticisms.
David Alves' research at a DOE facility is 
supported by an Associated Western Universities
Laboratory Graduate Fellowship.  
Work at LLNL performed under the auspices of the 
USDOE contract no.~W-7405-ENG-48.  
David Alves thanks his advisors, Kem Cook
and Bob Becker, for their
continued support.  
Ata Sarajedini was supported by the
National Aeronautics and Space Administration (NASA)
grant number HF-01077.01-94A from
the Space Telescope Science
Institute, which is operated by the Association of Universities for
Research in Astronomy, Inc., under NASA contract NAS5-26555.
Ata Sarajedini also wishes to thank Lick Observatory for their 
generosity and hospitality during his visit.
This collaboration was initiated at the 191st meeting 
of the American Astronomical 
Society held in Washington D.C.

\begin{figure}
\caption{Theoretical data showing the dependence of the absolute visual
magnitude of the red horizontal branch, $M_{V,HB}$ as a function of 
$\log(Age)$, where in age is in units of years.   Open circles connected with solid
lines are data from Vassiliadis and Wood (1993), for Y = 0.25, classical
mixing and Z = 0.008, 0.004, and 0.001, decreasing in luminosity in that
order.  Open triangles connected by a dashed line are data from
Seidel, Demarque, and Weinberg (1987) for Y = 0.25 and Z = 0.01.
Open squares connected by a dotted line are data from Lattanzio (1991
for Y = 0.25 and Z = 0.001.
The filled squares connected
by short dashed lines are from Sarajedini, Lee, and Lee (1995)
for [Fe/H] = $-1.7$ and $-0.7$ dex, where the former is brighter.
The solid line in the lower right corner of the figure (labeled) represents 
the dependence of $M_{V,HB}$ on age adopted in this paper.  The 
zero-point is arbitrary, and will be empirically calibrated.}
\end{figure}

\begin{figure}
\caption{Theoretical data showing the dependence of 
RGB-bump luminosity, $\log(L)$, and model mass, both
in solar units.  Filled triangles connected by
solid lines are data from Alongi et al. (1993) for Z = 0.008, Y = 0.25, and
classical and overshoot mixing schemes (brighter and fainter respectively).
Open squares connected by short-dashed lines are data from Sweigart and Gross
(1978) for Z = 0.01, classical mixing, and Y = 0.30 and 0.20 (brighter
and fainter respectively).  Filled circles connected with solid
lines are data from 
Vassiliadis and Wood (1993), for Y = 0.25, classical mixing and
Z = 0.008, 0.004, and 0.001, increasing in luminosity in that
order.  The long-dashed line is the mass dependence given by Fusi~Pecci
et al.~(1990), with an arbitrary $\log(L)$ zero-point.  The solid
line in the lower right corner of the figure (labeled) represents the dependence
of RGB-bump luminosity on mass adopted in this paper.  The
zero-point is arbitrary, and will be empirically calibrated.}
\end{figure}

\begin{figure}
\caption{Theoretical data for the luminosity of the AGB-bump as a function
of mass and metallicity.  
Top panel shows $\log(L)$ as function of $M$ (solar units)
for Z = 0.001 data; open squares from Castellani et al.~(1991) and filled
circles from Vassiliadis and Wood (1993).  Bottom panel shows $\log(L)$ as
a function of [Fe/H], for $M$ = 1.50 (filled circles) and $M$ = 0.85
(open squares), same symbols as above.  Projected lines from our analytic
fit to the data are shown as solid lines.  In the top panel, projections
at constant metallicities of [Fe/H] = $-1.7$, $-1.3$, and $-0.4$ dex are
plotted and labeled.  In the bottom panel, projections at constant mass
are shown and labeled ($M$ = 0.85, 0.95, 1.10, 1.30, 1.50).}
\end{figure}

\begin{figure}
\caption{Summary of theoretical data for the
absolute visual magnitudes of the horizontal branch ($M_{V,HB}$, filled squares),
the RGB-bump ($M_{V,RB}$, open circles), and the AGB-bump ($M_{V,AB}$, open
triangles) plotted as a function of $\log(Age)$, where age is units of years.
Each panel represents a different metallicity, [Fe/H], and is labeled.
See also Table~2.}
\end{figure}

\begin{figure}
\caption{Color-magnitude diagram of the SMC cluster NGC~411, derived
from archival {\it Hubble Space Telescope} WFPC2 data (PC-chip only). 
Arrow at V = 19.43 marks the red HB clump mean magnitude.  Arrow at
V = 18.74 is the giant branch bump.  See text for details of measuring
these values.}
\end{figure}

\begin{figure}
\caption{$\Delta$V$^{HB}_{Bump}$--[Fe/H] diagram, which shows metallicity
versus brightness difference between an RGB-bump or AGB-bump and the HB.
The filled squares are RGB-bump observational data for
Galactic globular clusters (see Table~3), 
and the solid
line labeled ``12 Gyr'' is the regression through these points.
The solid lines show the expected age-shift of $\Delta$V$^{HB}_{Bump}$
for the RGB-bump for ages of 10, 5, and 2 Gyr (labeled), 
assuming the brightness of the HB has no dependence on age, and an age of 12 Gyr
for the Galactic globular clusters.   The short-dashed lines show the
$\Delta$V$^{HB}_{Bump}$ age-shifts assuming $M_{V,HB}$ depends on age
for the RGB-bump for 10, 5, and 2 Gyr (see text).  The dotted lines
show the location of AGB-bumps in this diagram for ages of 12, 10, 5 and
2 Gyr, assuming the $M_{V,HB}$ depends on age.  Filled circles are observational data
for 4 Galactic globular cluster AGB-bumps (see text).  Observational data for 8 SMC
clusters shown as open circles with error bars.}
\end{figure}

\begin{figure}
\caption{$\delta\Delta$V$^{HB}_{Bump}$, the distance from the ``12 Gyr''
line in Figure~5 (Eqn.~7) in magnitudes, as a function
of age, in units of Gyr.  SMC cluster observational data shown as open circles with
error bars.  Below each SMC cluster point, we have labeled the metallicity.
RGB-bump observational data for Galactic globular clusters shown as
filled squares, error bars omitted for clarity (see Table~3).  
$\delta\Delta$V$^{HB}_{Bump}$ = 0 shown as a solid line.  Dotted
lines show our theoretical predictions for [Fe/H] = $-1.7$, $-1.3$,
$-0.7$, and $-0.4$ dex, where the latter three lines are very
similar.  The dashed line shows the approximations to our theoretical
predictions, given by Equations 6 \& 10 (see text).  The normalization
of the theoretical lines to $\delta\Delta$V$^{HB}_{Bump}$ = 0 at 12 Gyr
is also illustrated here.}
\end{figure}

\clearpage
\tighten

\begin{deluxetable}{llll}
\tablewidth{0pt}
\tablecaption{Reference RGB Coefficients}
\tablehead{
\colhead {[Fe/H]} &
\colhead {$\alpha$} & \colhead {$\beta$}  & \colhead {$\gamma$}
}
\startdata
$-1.7$  & $-484.46$ & 288.07 & $-42.31$ \nl
$-1.3$  & $-505.79$ & 295.19 & $-42.72$ \nl
$-0.7$  & $-625.90$ & 361.71 & $-51.95$ \nl
$-0.4$  & $-664.94$ & 382.99 & $-54.87$ \nl
\enddata
\end{deluxetable}

\clearpage

\begin{deluxetable}{lllrrr}
\footnotesize
\tablewidth{0pt}
\tablecaption{Theoretical Data}
\tablehead{
\colhead {[Fe/H]} & \colhead {Age (Gyr)} &
\colhead {$M$ ($M_{\odot}$)}  &
\colhead {$M_{V,HB}$} &
\colhead {$M_{V,RB}$} &
\colhead {$M_{V,AB}$}
}
\startdata
$-1.7$ & 12 &  0.865 & 0.445 &  0.185     & $-0.626$ \nl
       & 10 &  0.963 & 0.405 &  $-0.167$  & $-0.666$ \nl
       & 5  &  1.175 & 0.255 &  $-0.737$  & $-0.919$ \nl
       & 2  &  1.586 & 0.055 &  $-1.318$  & $-1.402$  \nl
$-1.3$ & 12 &  0.865 & 0.505 &  0.525     & $-0.460$ \nl
       & 10 &  0.963 & 0.465 &  0.335     & $-0.521$ \nl
       & 5  &  1.175 & 0.315 &  $-0.064$  & $-0.791$ \nl
       & 2  &  1.586 & 0.115 &  $-0.782$  & $-1.291$ \nl
$-0.7$ & 12 &  0.920 & 0.595 &  1.035     & $-0.342$ \nl
       & 10 &  1.024 & 0.555 &  0.835     & $-0.391$ \nl
       & 5  &  1.247 & 0.405 &  0.423     & $-0.600$ \nl
       & 2  &  1.680 & 0.205 &  $-0.351$  & $-0.969$ \nl
$-0.4$ & 12 &  0.947 & 0.640 &  1.290     & $-0.181$ \nl
       & 10 &  1.088 & 0.600 &  1.088     & $-0.229$ \nl
       & 5  &  1.284 & 0.450 &  0.662     & $-0.427$ \nl
       & 2  &  1.728 & 0.250 &  $-0.063$   & $-0.767$ \nl
\enddata
\end{deluxetable}

\clearpage
\begin{deluxetable}{lccccc}
\footnotesize
\tablenum{3}
\tablecolumns{5}
\tablewidth{0pt}
\tablecaption{Observational Data}
\tablehead{
\colhead{Cluster}  & \colhead{$V(HB)$} & \colhead{$V(Bump)$} &
\colhead{[Fe/H]} & \colhead{Age (Gyr)}
}
\startdata
  &  & SMC\tablenotemark{\ A} &  & \cr
NGC 411      & $19.43\pm0.01$ & $18.74\pm0.02$ & $-0.68\pm0.07$ & $1.4\pm0.2$ \cr
Kron 3       & $19.45\pm0.01$ & $19.38\pm0.04$ & $-1.16\pm0.09$ & $6.0\pm1.3$ \cr
Lindsay 113  & $19.15\pm0.01$ & $18.99\pm0.05$ & $-1.24\pm0.11$ & $5.3\pm1.3$ \cr
Lindsay 1    & $19.34\pm0.01$ & $19.32\pm0.03$ & $-1.35\pm0.08$ & $9.0\pm1.0$ \cr
NGC 416      & $19.74\pm0.01$ & $19.68\pm0.03$ & $-1.44\pm0.12$ & $6.9\pm1.1$ \cr
NGC 361      & $19.53\pm0.01$ & $19.48\pm0.05$ & $-1.45\pm0.11$ & $8.1\pm1.2$ \cr
NGC 339      & $19.46\pm0.01$ & $19.19\pm0.03$ & $-1.50\pm0.14$ & $6.3\pm1.3$ \cr
NGC 121      & $19.73\pm0.01$ & $19.51\pm0.02$ & $-1.71\pm0.10$ & $11.9\pm1.3$ \cr
  &  & Galaxy\tablenotemark{\ A} &  & \cr
NGC 5972      & $16.60\pm0.02$ & $17.32\pm0.04$ & $-0.30\pm0.09$ & $old$ \cr
NGC 6352      & $15.28\pm0.02$ & $15.86\pm0.06$ & $-0.51\pm0.08$ & $12.0\pm1.0$ \cr
NGC 6637      & $16.00\pm0.02$ & $16.53\pm0.07$ & $-0.59\pm0.19$ & $old$ \cr
NGC 104       & $14.10\pm0.15$ & $14.55\pm0.05$ & $-0.71\pm0.08$ & $11.6\pm1.0$ \cr
NGC 6171      & $15.75\pm0.05$ & $16.00\pm0.05$ & $-0.99\pm0.06$ & $15.0\pm1.8$ \cr
NGC 1261      & $16.70\pm0.10$ & $16.70\pm0.05$ & $-1.31\pm0.09$ & $13.1\pm1.1$ \cr
NGC 1851      & $16.15\pm0.02$ & $16.15\pm0.03$ & $-1.36\pm0.09$ & $11.6\pm0.8$ \cr
NGC 2808      & $16.25\pm0.10$ & $16.20\pm0.05$ & $-1.37\pm0.09$ & $14.1\pm1.3$ \cr
NGC 5904      & $15.11\pm0.05$ & $15.05\pm0.05$ & $-1.40\pm0.06$ & $14.0\pm1.0$ \cr
NGC 6752      & $13.75\pm0.15$ & $13.65\pm0.05$ & $-1.54\pm0.09$ & $17.0\pm1.8$ \cr
NGC 7006      & $18.72\pm0.10$ & $18.55\pm0.05$ & $-1.59\pm0.07$ & $13.5\pm1.1$ \cr
NGC 5272      & $15.65\pm0.05$ & $15.40\pm0.05$ & $-1.66\pm0.06$ & $13.6\pm0.8$ \cr
NGC 5897      & $16.35\pm0.15$ & $16.08\pm0.05$ & $-1.68\pm0.11$ & $14.6\pm1.8$ \cr
NGC 1904      & $16.25\pm0.10$ & $15.95\pm0.05$ & $-1.69\pm0.09$ & $14.1\pm1.3$ \cr
NGC 6397      & $13.00\pm0.10$ & $12.60\pm0.10$ & $-1.91\pm0.14$ & $17.8\pm1.7$ \cr
$MP$\tablenotemark{B} & $15.86\pm0.05$ & $15.35\pm0.05$ & $-2.15\pm0.08$ & $17.2\pm2.0$ \cr
\enddata
\tablenotetext{A}{References: Mighell et al.~(1998), Fusi Pecci et al.~(1990),
Sarajedini \& Forrester (1995), Chaboyer, Demarque, \& Sarajedini (1996), this paper.}
\tablenotetext{B}{MP is metal-poor, a co-added CMD of M15, M92, \& NGC~5466.}
\end{deluxetable}



\clearpage
\plotone{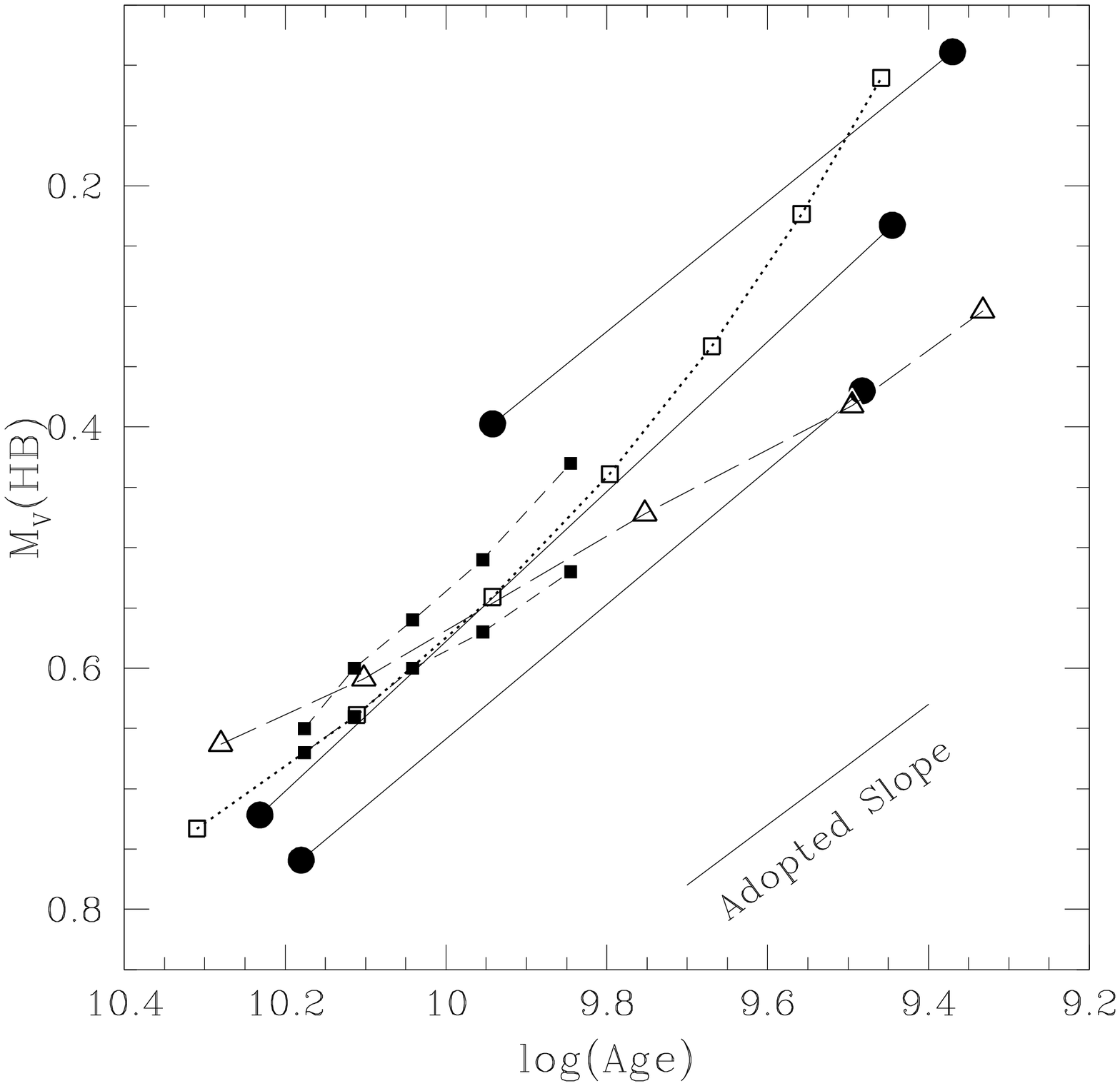}
\clearpage
\plotone{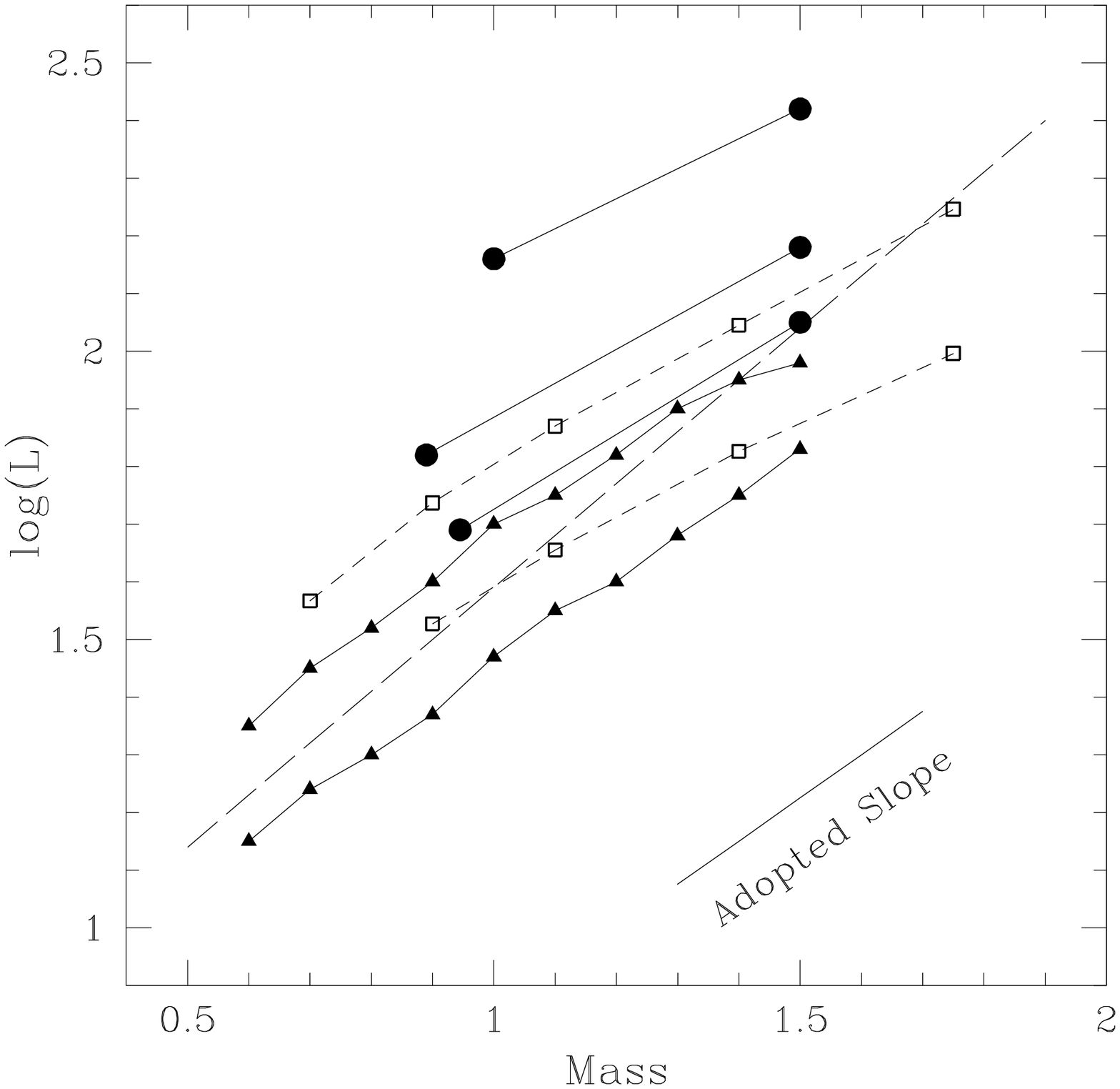}
\clearpage
\plotone{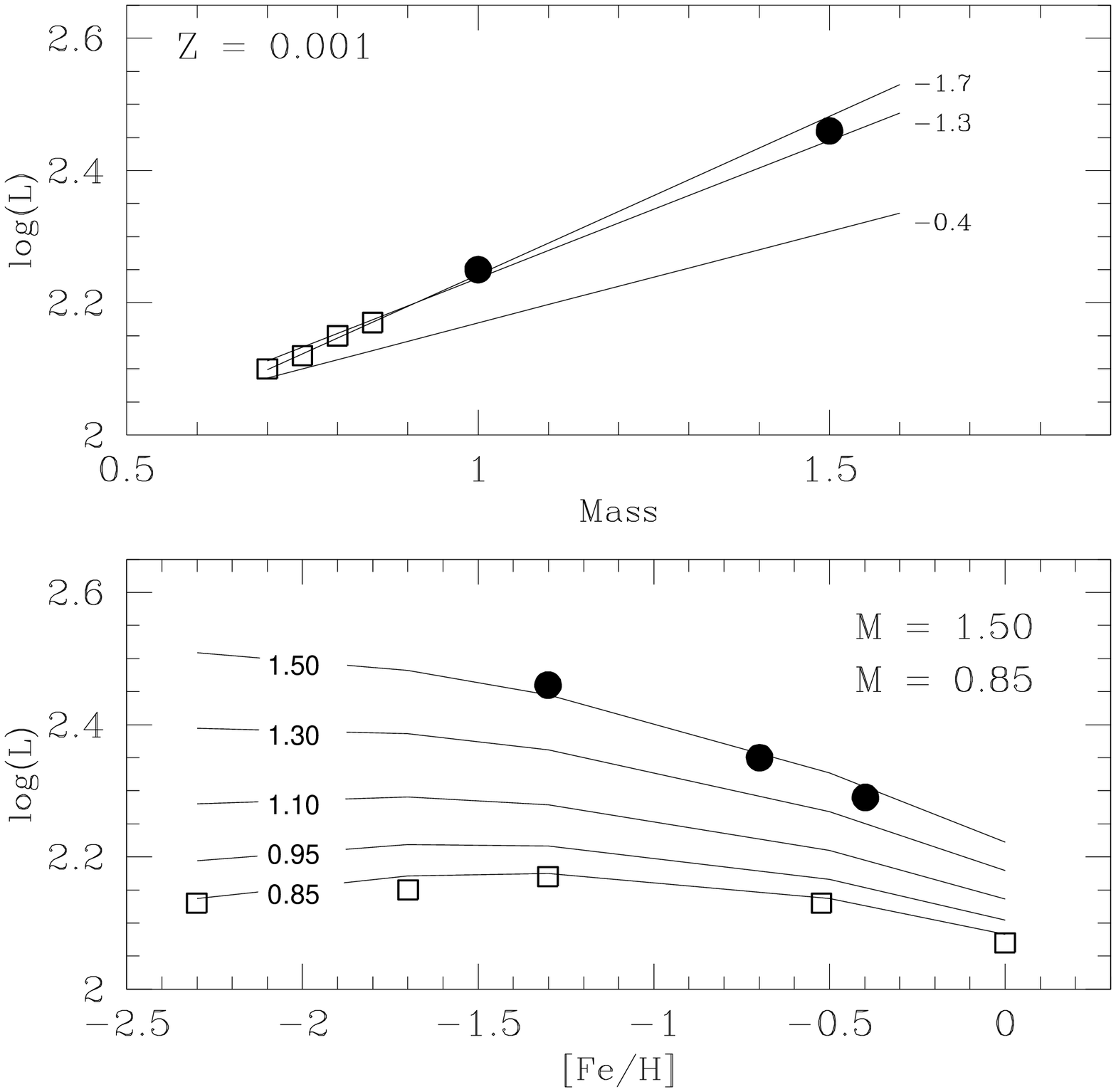}
\clearpage
\plotone{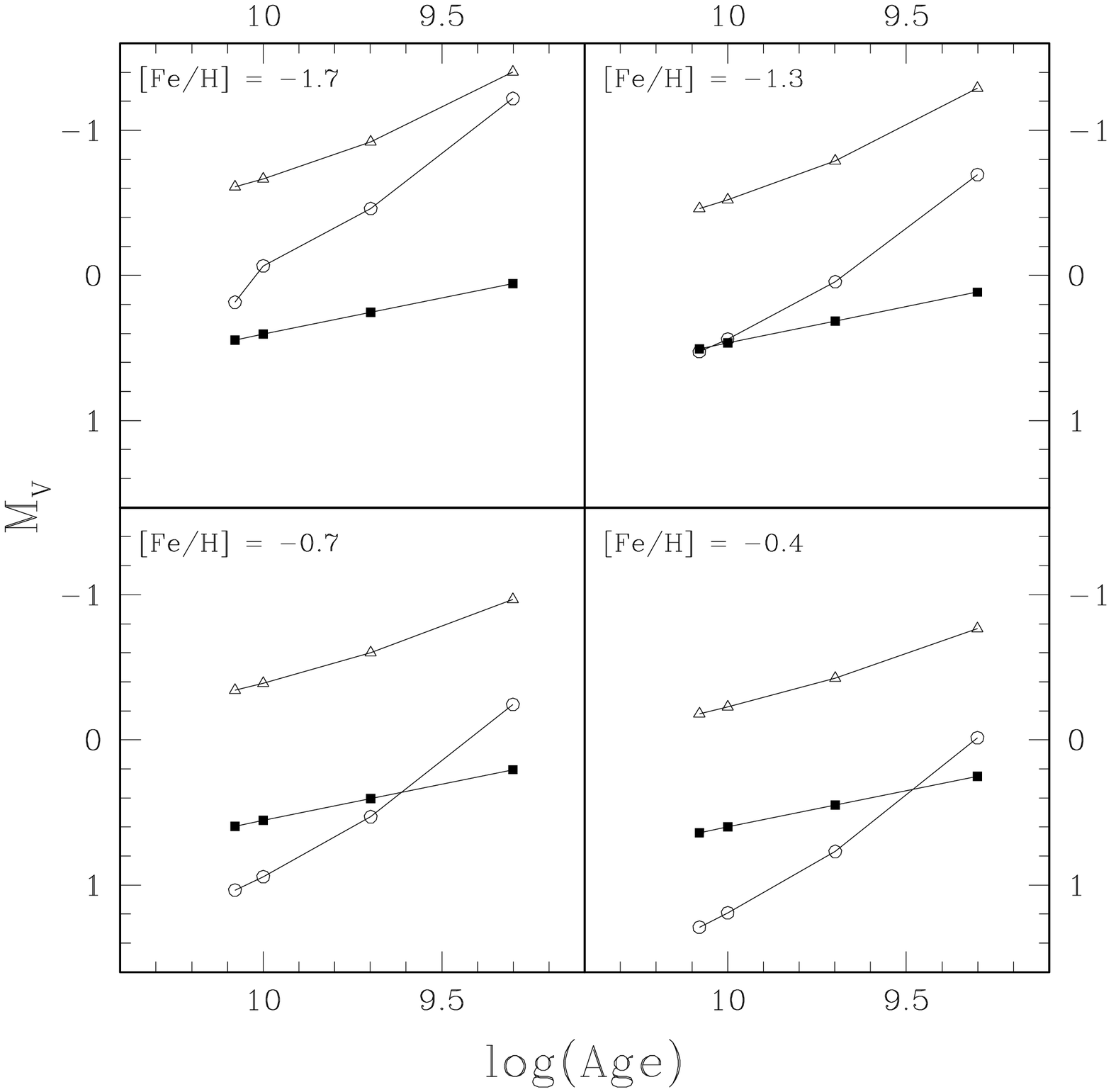}
\clearpage
\plotone{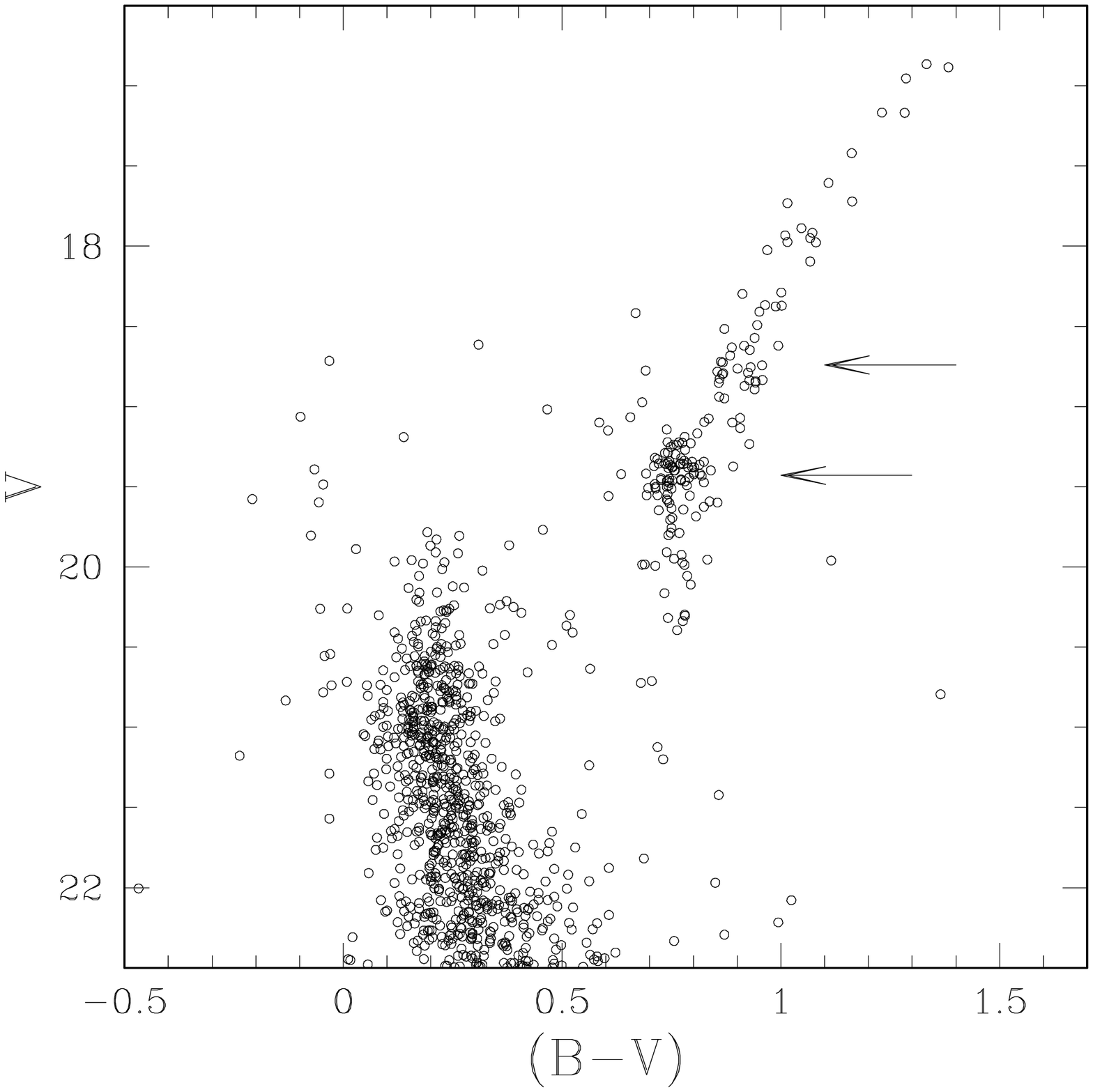}
\clearpage
\plotone{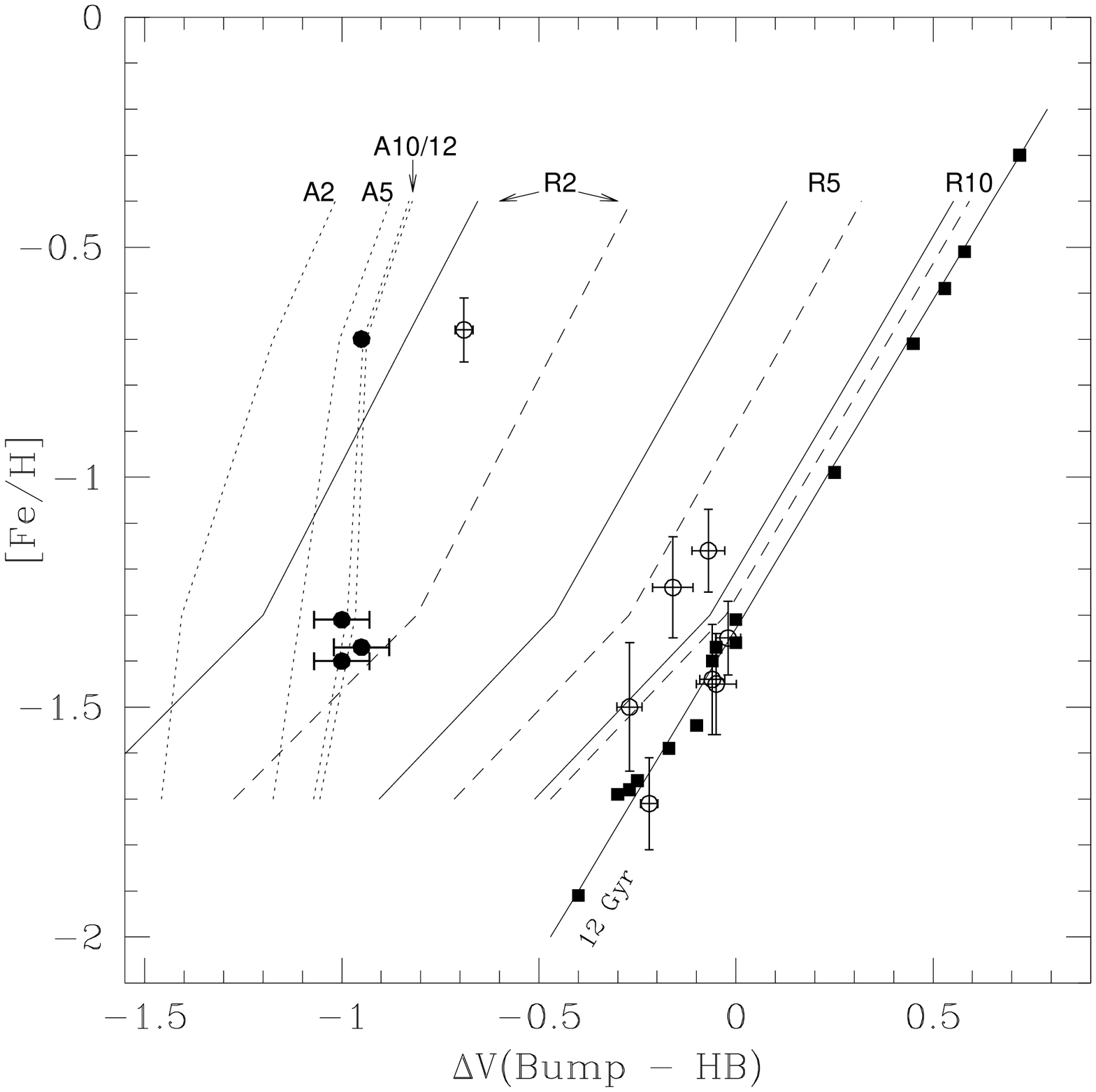}
\clearpage
\plotone{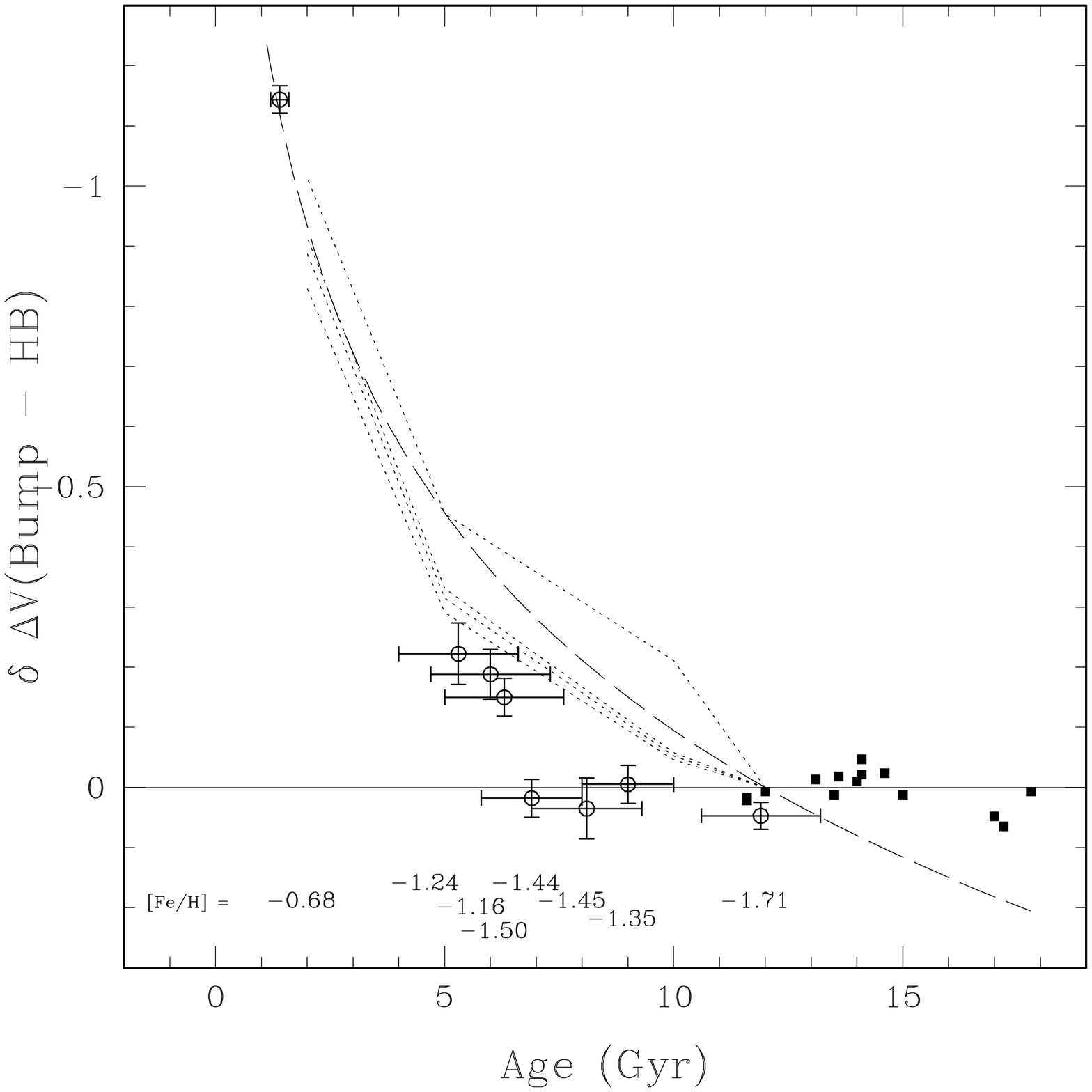}



\clearpage
\begin{references}


\reference{}{ Alcock et al. {\tt(MACHO Collaboration)} 1997 \apj, 482, 89 }
\reference{}{ Alcock et al. {\tt(MACHO Collaboration)} 1998 {\it in preparation} }
\reference{}{ Alongi M., Bertelli, G., Bressan, A., Chiosi, C., Fagaotto, F.,
Greggio, L., \& Nasi, E. 1993 \aaps, 97, 851 }
\reference{}{ Alves, D. 1998 Ph.D dissertation, {\it in preparation} }
\reference{}{ Alves, D. et al. {\tt(MACHO Collaboration)} 1998 BAAS 191, 115.01 }
\reference{}{ Bertelli, G., Bressan, A., Chiosi, C., Fagotto, F., \& Nasi, E. 
1994, \aaps, 106, 275 }
\reference{}{ Bono, G. \& Castellani, V. 1992 \aap, 258, 385 }
\reference{}{ Bono, G., Castallani, V., Degl'Innocenti, S., \& Pulone, L. 1995 \aap, 297, 115}
\reference{}{ Cannon, R. 1970 \mnras, 150, 111 }
\reference{}{ Caputo, F., Castellani, V., Chieffi, A., Pulone, L., \& Tornambe, A. 1989
\apj, 340, 241 }
\reference{}{ Carney, B., Storm, J., \& Jones, R. 1992 \apj, 386, 663 }
\reference{}{ Cassisi, S. \& Salaris, M. 1997 \mnras, 285, 593 }
\reference{}{ Castellani, V., Chieffi, A., Pulone, L., \& Tornambe, A. 1985 \aap, 296, 204 }
\reference{}{ Castellani, V., Chieffi, A., \& Norci, L. 1989 \aap, 216, 62 }
\reference{}{ Castellani, V., Chieffi, A., \& Pulone, L. 1991 \apjs, 76, 911 }
\reference{}{ Catelan, M. \& De Freitas Pacheco \pasp, 1996, 108, 166 }
\reference{}{ Chaboyer, B., Demarque, P., \& Sarajedini, A. 1996 \apj, 459, 558 }
\reference{}{ Da Costa, G. S., \& Mould, J. R. 1986, \apj, 305, 214 }
\reference{}{ Ferraro, F. 1992 {\it Mem.S.A.It.} 63, 491 }
\reference{}{ Fusi Pecci, F., Ferraro, F., Crocker, D., \& Buonanno, R. 
1990 \aap, 238, 95 }
\reference{}{ Gallart, C. 1998 \apjl, 495, 43 }
\reference{}{ Hesser, J., Harris, W., Vandenbergh, D., Allwright, J. Shott, P,
\& Stetson, P. 1987 \pasp, 99, 739}
\reference{}{ Hesser J., et al. 1996 {\it Journal Korean Ast. Soc.}, 29, 1 }
\reference{}{ Iben, I. 1968 Nature, 220, 143 }
\reference{}{ Iben, I. \& Laughlin, G. 1989 \apj, 341, 312 }
\reference{}{ King, C., Da Costa, G., \& Demarque, P. 1985 \apj, 299, 674}
\reference{}{ Lattanzio, J. 1986 \apj, 311, 708 }
\reference{}{ Mengel, J,. Demarque, P., Sweigart, A., \& Gross, P. 1979 \apjs, 40, 733}
\reference{}{ Mighell, K. J., Sarajedini, A., \& French, R. S. 1998, {\it in preparation} }
\reference{}{ Noriega-Mendoza, H. \& Ruelas-Mayorga, A. 1997, \aj, 113, 722 }
\reference{}{ Pulone, L. 1992 {\it Mem.S.A.It.} 63, 485 }
\reference{}{ Refsdale, S. \& Weigart, A. 1970 \aap, 6, 426}
\reference{}{ Renzini, A. \& Fusi Pecci, F. 1988 \araa, 26, 199 }
\reference{}{ Sarajedini, A. 1994, \aj, 107, 618 }
\reference{}{ Sarajedini, A. \& Forrester, W. 1995 \aj, 109, 1112 }
\reference{}{ Sarajedini, A., Lee, Y., \& Lee, D. 1995 \apj, 450, 712}
\reference{}{ Sarajedini, A., Chaboyer, B., \& Demarque, P. 1997 \pasp, 109, 1321}
\reference{}{ Sarajedini, A. 1998 AJ, {\it in press} }
\reference{}{ Seidel, E., Demarque, P., \& Weinberg, D. 1987 \apjs, 63, 917 }
\reference{}{ Smecker-Hane, T., Stetson P., Hesser, J., \& Lehnert, M. 1994 \aj, 108, 507}
\reference{}{ Straniero, O. \& Chieffi, A. 1991 \apjs, 76, 525 }
\reference{}{ Sweigart, A. \& Gross, P. 1978 \apjs, 36, 405 } 
\reference{}{ Sweigart, A., Greggio, L., \& Renzini, A. 1989 \apjs, 69, 911 } 
\reference{}{ Sweigart, A., Greggio, L., \& Renzini, A. 1990 \apj, 364, 527 } 
\reference{}{ Thomas, H. 1967 {\it Zeit Astrophyzik}, 67, 420 }
\reference{}{ Vassiliadis, E. \& Wood, P. 1993 \apj, 413, 641 }
\reference{}{ Walker, A. 1992 \apjl, 390, 81 }

\end{references}
\end{document}